\documentclass[openbib]{article}
\usepackage{a4,graphicx,epsf,multicol}
\newtheorem{fig}{Fig.}
\begin{document}
\bibliographystyle{prlprb}
\begin{center}
{\large\bf Proximity Effect, Andreev Reflections, and Charge Transport in Mesoscopic Superconducting--Semiconducting Heterostructures}\\[1em]
A. Jacobs and R. K\"ummel\\
{\em Institut f\"ur Theoretische Physik der Universit\"at W\"urzburg, Am Hubland, D-97074 W\"urzburg, Germany}\\[1em]
H. Plehn\\
{\em Rechenzentrum der Universit\"at W\"urzburg, Am Hubland, D-97074 W\"urzburg, Germany}\\[1em]
Received (\hspace{27mm})\\[2em]
\begin{minipage}{14cm}
In the quasi--twodimensional (Q2D) electron gas of an InAs channel between an AlSb substrate and superconducting Niobium layers the proximity effect induces a pair potential so that a Q2D mesoscopic superconducting--normal--superconducting (SNS) junction forms in the channel. The pair potential is calculated with quasiclassical Green's functions in the clean limit. For such a junction alternating Josephson currents and current--voltage characteristics (CVCs) are computed, using the non--equilibrium quasiparticle wavefunctions which solve the time--dependent Bogoliubov--de Gennes Equations. The CVCs exhibit features found experimentally by the Kroemer group: A steep rise of the current at small voltages (``foot'') changes at a ``corner current'' to a much slower increase of current with higher voltages, and the zero--bias differential resistance increases with temperature. Phase--coherent multiple Andreev reflections and the associated Cooper pair transfers are the physical mechanisms responsible for the oscillating Josephson currents and the CVCs. Additional experimental findings not reproduced by the theory require model improvements, especially a consideration of the external current leads which should give rise to hybrid quasiparticle/collective--mode excitations.
\end{minipage}
\end{center}

\vspace*{0.5em}
\begin{multicols}{2}
\centerline{\bf 1. Introduction}\vspace{1em}
The scattering of electrons into holes and vice versa by spatial variations of the superconducting pair potential, which occurs at the interfaces between normal (N) and superconducting (S) regions \cite{and64,and66,mcm68}, influences significantly the electronic structure and the transport properties of mesoscopic superconducting--semiconducting--superconducting (SSmS) heterostructures. (In this context ``electrons'' and ``holes'' are quasiparticle excitations with momenta above and below the Fermi surface of the degenerate semiconductor.) This scattering, also called Andreev reflection, establishes phase--coherence between the superconducting regions and transfers Cooper pairs across the semiconducting layer(s), thus giving rise to Josephson currents.

Most experiments on SSmS junctions involve a quasi--twodimensional (Q2D) electron gas in a narrow channel between the superconducting banks. The electrons are either in the inversion layer which forms between p--InAs and Niobium (Nb) \cite{tak85,chr96,chr97}, or they accumulate in MBE--grown quantum wells formed by n--type InAs between a p--type InAs buffer layer \cite{nit92}, see also \cite{nit94}, or an AlSb substrate and two \cite{nit92,kro94,kro94a,bas98} or up to 300 \cite{kro94a,tho98} Nb electrodes, separated by insulating AlSb layers. --- Transport properties of SSmS and SNS junctions involving high--temperature superconductors are investigated, too \cite{pet97a,pet97b}.

One observes peculiar (subgap) structures \cite{chr96}--\cite{nit92},\linebreak \cite{kro94}--\cite{tho98} and spikes \cite{bas98} in the differential resistances, and current-voltage characteristics (CVCs) show first a steep rise of the current, then a plateau and finally a linear current increase with ``high'' voltages \cite{kro94a,tho98}. The observed phenomena are interpreted in terms of Andreev reflections \cite{chr96}--\cite{tho98}. Nitta et al. \cite{nit92,nit94} assume that the spatially varying pair potential, required for these reflections, is induced in the InAs by the proximity effect. Chrestin et al. \cite{chr96,chr97} extend the essentially one\-dimensional, dirty--limit model of Aminov et al. \cite{ami96} to the Q2D case and obtain a gap in the density of states which is understood to originate from the finite average lifetime of quasiparticles. Volkov et al. \cite{vol95} find a small gap in the excitation spectrum of the Q2D electron gas, too; multiplying it by the phase difference between the superconducting banks they consider it as the effective order parameter. 

In this paper we compute the pair potential (order parameter) $\Delta$, induced in the InAs below the superconducting Nb electrodes by the proximity effect, from the self--consistency equation (\ref{Selbstkonsistenzgleichung}). The wavefunctions of quasiparticles in the Q2D electron gas, Andreev reflected from the walls of the  superconducting pair potential well, are calculated from the time--dependent Bogoliubov--de Gennes equations. We assume a potential drop $eV$ across the Sm region and a model in which the wavefunctions in the Sm region evolve from wavefunctions at energies $E \geq \Delta$ in the S banks. With these wavefunctions alternating Josephson currents are computed, and their time--averages yield the current--voltage characteristics. Comparison with experiments and a critique of the model terminate the paper.

\vspace{1em}
\centerline{\bf 2. Proximity Effect}\vspace{1em}

For the numerical computation of the proximity--induced pair potential and the current--voltage characteristics we use system parameters similar to the ones reported by the Kroemer group \cite{kro94,kro94a,tho98}. We assume that the InAs channel extends in x--direction over 15 nm (thickness $L_x$), and in z--direction the extensions of the insulating layers between the Nb electrodes, $2a$, and of the Nb electrodes themselves, $D-a$, are about 500\nolinebreak nm. The width $L_y$ of the sample in y--direction is of the order of magnitude of 100 $\mu$m. Furthermore, in the InAs channel the ``parabolic--equivalent'' effective mass of the electrons is approximately $m^*=0.05\, m_0$ ($m_0$: free electron mass) \cite{kro98pc}.

We model the electronic structure of these systems in the following way. In the Q2D electron gas, below the superconducting Nb electrodes, the proximity effect, i.e. the diffusion of Cooper pairs, induces a finite superconducting pair potential $\Delta (z)$. In the InAs channel, according to Fig. \ref{pp}a, the variation of the pair potential along the x--axis between the Nb electrode at $x=0$ and the substrate at $x= L_x=-0.08\, \xi_0$ may be neglected.\linebreak ($\xi_0$ = 190 nm is the BCS coherence length of bulk Nb.)

\vspace{-1em}
\begin{center}
\includegraphics[width=8cm]{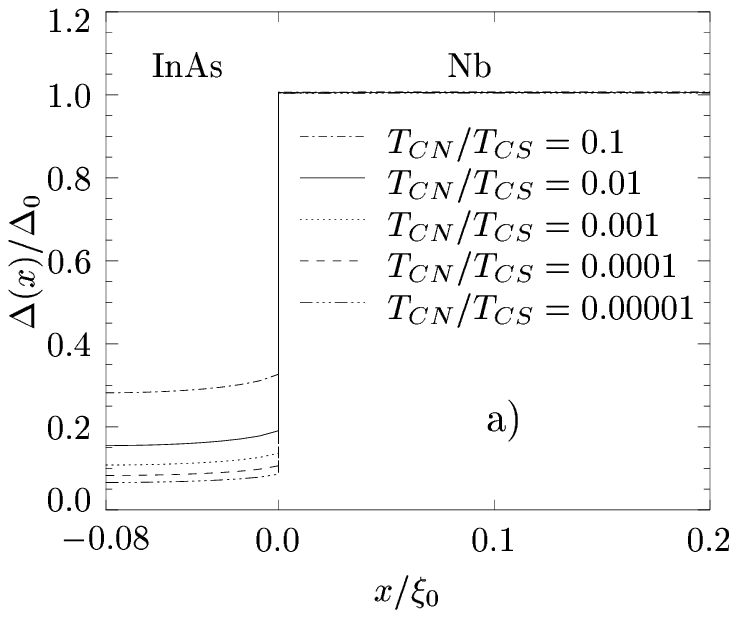}\\[-7mm]
\includegraphics[width=8cm]{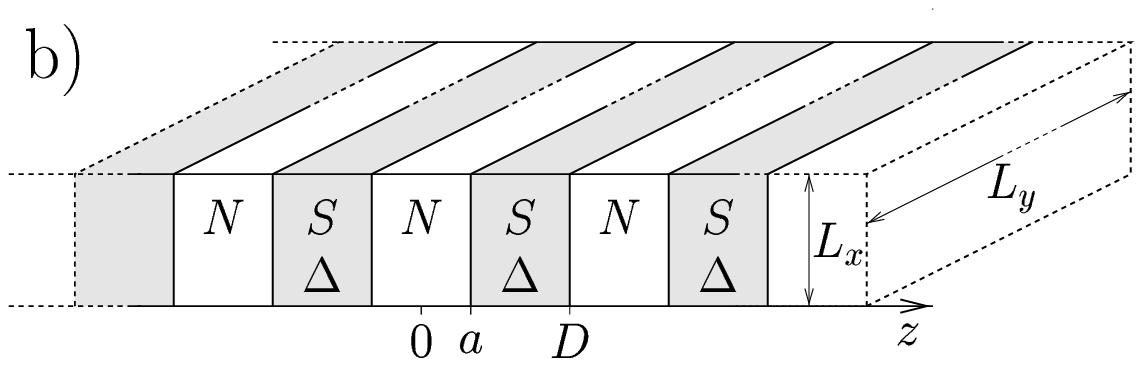}
\end{center}
\vspace{-2em}
\begin{fig}{\bf a)} The pair potential in a superconducting Nb electrode and the proximity--induced pair potential\linebreak in the InAs channel of the samples investigated by the Kroemer group \cite{kro94a,tho98}. The critical temperature of Nb is $T_{CS}$ = 9.2 K and that of InAs is $T_{CN}$, see text.\linebreak {\bf b)} The (Q2D) SNS--superlattice as a model of the channel structure in \cite{tho98}. \label{pp}
\end{fig}

The proximity--induced pair potential in the clean limit is calculated with quasiclassical Green's functions $\hat{g}(\vec{e}_k,x,\omega_n)$ and the self--consistency equation 
\begin{equation}
\label{Selbstkonsistenzgleichung}
\Delta(x) = \frac{k_{B}T\sum_{n} \int d\Omega_{k}(4\pi)^{-1} \mbox{Tr}[\hat{g}(\vec{e}_k,x,\omega_{n})(\hat{\tau}_{1} -i\hat{\tau}_{2})]}{\mbox{ln}(T/T_{C}(x)) + \sum_{n} 1/(n-0.5)}
\end{equation}
with the $x$--dependent critical temperature
\begin{equation}
T_{C}(x) = \left\{ \begin{array}{cl}T_{CN} & \mbox{in N} \\ T_{CS} & \mbox{in S}\end{array}\right. .
\end{equation}
The solid angle integration is over all momentum directions $\vec{e}_k$. The summation $\Sigma_{n}$ over the Matsubara frequencies $\omega_n=(2n-1)\pi k_BT/\hbar$ can be cut off at $\hbar\omega_{n}=\hbar\omega_{c}\approx 10\, k_{B}T_{C}$ \cite{kie87,bru90}. $\hat{\tau}_i$ are the Pauli matrices. [Eq. (\ref{Selbstkonsistenzgleichung}) results from a combination of Kieselmann's \cite{kie87} eqs. (2.6c) and (2.7b). It has been extended by Bruder \cite{bru90} to the case of anisotropic pairing interactions.] More details are given in Ref. \cite{ple94}. We choose the temperature\linebreak $T=2.2$ K and use the following material parameters:\linebreak 0.21 eV = Fermi energy of InAs, 5.32 eV = Fermi energy of Nb, 0.2 = ratio of the Fermi velocity in InAs to that in Nb. The temperature--parameter ratio $T_{CN}/T_{CS}$ measures the pairing inter\-action strength in InAs relative to that of Nb. This parameter is unknown, but we see that even for ratios as small as 0.001 one gets a pair potential $\Delta$ in the\linebreak InAs which is about 10\% of the Nb pair potential\linebreak $\Delta_0=1.5$ meV. Therefore, {\em in the InAs layer} superconducting and semiconducting regions should alternate in z--direction, along which charge transport occurs. As a consequence the electronic structure is similar to that of ballistic SNS junctions in which only the spatial variation of the pair potential breaks translational invariance in z--direction, see Fig. 1b.


\vspace{1em}
\centerline{\bf 3. Quasiparticle--Dynamics}\vspace{1em}

\noindent{\bf The Model}\vspace{1em}

We use the formalism of the time--dependent Bogoliubov--de Gennes Equations (TdBdGE) \cite{kue69} in order to describe the charge transport in the super--semi--superconducting junction formed by the InAs channel with the proximity induced pair potential of magnitude $\Delta$. We treat the system as a quasi--twodimensional SNS junction. We assume that before times $t=t_0$ the system is in equilibrium, and each quasiparticle state is characterized by a set of quantum numbers $k$. At times $t \ge t_0$ there appears a potential difference $eV$ across the semiconducting region of length $2a$ between the superconducting pair potentials. It may be due either to a voltage $V$ switched on at $t=t_0$ (voltage bias), or to an applied current which increases above the criti\-cal Josephson current at $t=t_0$ (current bias). The quasiparticle--wavefunction 
\begin{equation}
\label{PSIK}
\Psi_k(\vec{r},t)={u_k(\vec{r},t)\choose v_k(\vec{r},t)}
\end{equation}
with electron component $u_k(\vec{r},t)$ and hole component $v_k(\vec{r},t)$, which evolves from the equilibrium state $k$ under the influence of the electric field $- \vec{e}_zV/2a$\ in the Sm region, is determined by the TdBdGE 
\begin{equation}
\label{tdbdg}
i\hbar\frac{\partial}{\partial t}\Psi_k(\vec{r},t)=\mathcal{H}(\vec{r},t)\Psi_k(\vec{r},t)
\end{equation}
where
\begin{equation}
\label{hcal}
\mathcal{H}(\vec{r},t)=\left(\begin{array}{cc}H_0(\vec{r},t) & \Delta(z)\\ \Delta(z) & -H_0^*(\vec{r},t)\end{array}\right)
\end{equation}
and
\begin{equation}
\label{hnull}
H_0(\vec{r},t)=\frac{1}{2m^*}\left[\frac{\hbar}{i}\vec{\nabla}+\vec{e}_z\frac{\hbar}{4a}\Phi(t)\right]^2+U(x)-\mu .
\end{equation}
%
%
%
%
We model \cite{kue90} the spatial variations of the pair potential $\Delta(z)$ and of the phase difference $\Phi(t)$ (which satisfies the Josephson equation) by step--functions $\Theta(\zeta)$ (equal to 1, if $\zeta\ge 0$, and zero otherwise), using a gauge, where
\begin{equation}
\label{del}
\Delta(z)=\Delta\cdot\Theta(|z|-a)\Theta(D-|z|)
\end{equation}
is real and 
%
%
%
%
\begin{equation}
\label{phit}
\Phi(t)=\left[\frac{2eV}{\hbar}(t-t_0)+\Phi_0\right] \Theta(a-|z|);\quad e=+|e|.
\end{equation}
$\Phi_0 \equiv \Phi(t_0)$, in the case of current bias, is the phase difference at which the applied current is equal to the criti\-cal Josephson current, whereas in the case of a voltage--biased junction $\Phi_0=0$. The scalar potential $U(x)$ describes the quantum well which confines the InAs electrons in x--direction. In an admittedly rather crude way we model this confinement by demanding that the quasiparticle--wavefunctions vanish in $ x = 0$ and $x = L_x$. This means that we assume specular reflection of the electrons not only at the band--edge jump (of about\linebreak 1.4 eV) from InAs to AlSb but also at the band--edge drop (of about 4.9 eV) to the bottom of the Niobium conduction band. We assume translational invariance in y--direction. The chemical potential $\mu$ is that of the energy and particle reservoir, to which the channel is coupled. We neglect all influences of entropy production on the chemical potential, because the number of degrees of freedom of the reservoir is assumed to be very much larger than that of the channel. Then $\mu$ is the same as in the equilibrium situation and equal to the Fermi energy of the InAs channel.

We expect that the current--voltage characteristics  calculated for the super--semi--superconducting junction modeled by eqs. (\ref{tdbdg})--(\ref{phit}) will qualitatively show the same features as the ones of superlattices, which may be seen as many junctions in series.

According to the initial condition stated above the solutions of the TdBdGE (\ref{tdbdg}) must turn into the stationary quasiparticle--wavefunctions of an SNS junction in the limit of vanishing voltage, $V \rightarrow 0$. These wavefunctions correspond to the ones given in Ref. \cite{gun94}. There are two classes of stationary states: i) Scattering states with absolute value of energy, $|E|$, larger than the maximum value of the pair potential, $\Delta$, and ii) bound states with $|E|<\Delta$; E is measured relative to the chemical potential $\mu $. The existing theories of ballistic, dissipative charge transport in weak links differ in the contribution of the two types of voltage--dependent solutions which evolve from the two classes of stationary states. In Ref. \cite{gun94a} the solutions evolving from the scattering states play the dominant role, the ones evolving from the bound states matter only for very small voltages. (Refs. \cite{oct83,ave95,bratus97} and related theories consider only the evolution from scattering states.) In Ref. \cite{kue90}, on the other hand, current--voltage characteristics (with negative differential resistance observed in short metallic weak links used for millimeter--wave mixers \cite{mat95}) were calculated from solutions which evolve from the bound states. The physical difference between the two classes of theories consists in the assumption about the rate at which quasiparticles, in a sequence of relaxation cycles, start their motion in the electric field from energies $|E|<\Delta$: In Ref. \cite{gun94a} quasiparticles start after each relaxation cycle, whereas in Ref. \cite{kue90} they start after each Andreev reflection. The idea behind the assumption of Ref. \cite{kue90} is the following: If the SNS junction is connected to the current or voltage source by {\em normal conducting} external current leads, practically all quasiparticle excitations, forming the shifted Fermi sphere of the current carrying normal leads, have energies $|E|<\Delta$ so that they are totally Andreev reflected at the {\em external} interfaces between leads and junction. When decaying in the superconducting banks of the junction they form Cooper pairs and induce collective (supercurrent carrying) modes \cite{kue69,mat78,blo82}. 
%
%
%
%
These collective modes should turn again into quasiparticle excitations within the\linebreak N layer of the junction, with initial energies $|E|<\Delta$ as in the external leads. 
%
%
%
%
Because of charge conservation the rate at which the collective modes turn into quasiparticle excitations is equal to the rate of Andreev scattering at any of the interfaces.

The BCS theory of superconductivity and the related TdBdGE describe a superconductor as a grand canoni\-cal ensemble. If the Cooper pairs created (destroyed) by Andreev reflections in an SNS junction directly go to (come from) a reservoir of Cooper pairs, with no normal current leads between junction and reservoir, the model of Ref. \cite{gun94a} should apply. If, on the other hand, the reservoir is connected to the junction by normal conducting current leads, the model of Ref. \cite{kue90} should be appropriate. In the latter case one should analyze in detail the spatial distribution of the hybrid excitations, consisting of quasiparticles (in the normal current leads and the normal region of the junction) and collective supercurrent carrying modes (in the S layers). This is being investigated presently. In this paper we disregard, like Refs. \cite{gun94a}--\cite{bratus97}, the possible influence of normal current leads and assume that only quasiparticles with energies $|E|\ge\Delta$ are incident from the superconducting banks onto the NS interfaces of the junction.

\smallskip
\vspace{1.5em}
\noindent{\bf Multiple Andreev Reflections}\vspace{1.5em}

There are four different types of incident quasiparticles with energy $E\ge\Delta$. Their wavefunctions, which satisfy eq. (\ref{tdbdg}) in the superconducting banks and which are normalized to the volume $L_xL_y(D-a)$ of one S layer,\pagebreak\  are given by 
\begin{displaymath}
\Psi_{S,in}^{\alpha\beta}(E;\vec{r},t) = \eta(x,y)\cdot e^{-\frac{i}{\hbar}Et}N^\beta(E){1\choose\gamma(E)^\beta}
\end{displaymath}
\begin{equation}
\label{psisin}
\times e^{i\alpha\left[k_{zf}+\beta\kappa(E)\right]z}\Theta(-\alpha\beta z-a)\Theta(D-|z|)
\end{equation}
with
\begin{equation}
\label{eta}
\eta(x,y) = \sin\left(k_xx\right)e^{ik_yy}\; ;\quad k_x=\frac{\pi}{L_x}s,\; s=1,2\;\; ,
\end{equation}
describing the plane waves in y--direction and the standing waves in the quantum well forming the channel. At low temperatures, for $\mu=0.2$ eV, $L_x=15$ nm, and $m^*=0.053\, m_0$, only the subband states with $s=1$ and $s=2$ are occupied. The indices $\alpha$ and $\beta$ in eq. (\ref{psisin}) indicate the character of the quasiparticle: it is electron--like ($\beta=+1$) or hole--like ($\beta=-1$) with positive ($\alpha=+1$) or negative ($\alpha=-1$) momentum in z--direction. Thus, the complete set of quantum numbers characterizing the quasiparticle--wavefunction (\ref{PSIK}) is $k\equiv\{s;k_y;(E,\alpha,\beta)\}$.
\begin{equation}
\label{norm}
N^\beta(E) = \sqrt{\frac{2}{L_xL_y(D-a)\left[1+\gamma(E)^{2\beta}\right]}}\cdot\Theta(E-\Delta)
\end{equation}
is the normalization factor, with
\begin{equation}
\label{gamma}
\gamma(E)=\frac{E-\mbox{sign}(E)\sqrt{E^2-\Delta^2}}{\Delta} \quad \mbox{for }|E|\ge\Delta\;\; .
\end{equation}
For $|E|<\Delta$ 
\begin{equation}
\label{gammakleiner}
\gamma(E)=\frac{E-i\sqrt{\Delta^2-E^2}}{\Delta}\;\; .
\end{equation}
$\gamma(E)$ is the probability amplitude of Andreev reflection of a quasiparticle with energy $E$. [Arguments of $\gamma(E)$ less than $\Delta$ appear in the probability amplitudes of\linebreak {\em multiple} Andreev reflections $A_n^\alpha(E)$, see eq. (\ref{AE}).] The wave number of quasiparticle propagation in z--direction $\alpha\left[k_{zf}+\beta\kappa(E)\right]$ is determined by the component of the Fermi momentum perpendicular to the interfaces
\begin{equation}
\label{kzf}
k_{zf} = \sqrt{k_F^2-k_\varrho^2}
\end{equation}
and
\begin{equation}
\label{kappa}
\kappa(E)=\left\{\begin{array}{ll}\mbox{sign}(E)\frac{\sqrt{E^2-\Delta^2}}{\hbar v_{zf}} & \mbox{for }|E|\ge\Delta\\[2mm] i\frac{\sqrt{\Delta^2-E^2}}{\hbar v_{zf}} & \mbox{for }|E|<\Delta\end{array}\right.\;\; ,
\end{equation}
where $\frac{\hbar^2k_F^2}{2m^*}=\mu$, $\vec{k}_\varrho\equiv\vec{e}_xk_x+\vec{e}_yk_y$, and $v_{zf}=\frac{\hbar k_zf}{m^*}$.

The wavefunctions for the quasiparticles which are Andreev reflected at the NS interfaces and travel back in the S banks are:
\begin{displaymath}
\Psi_{S,out}^{\alpha\beta}(E;\vec{r},t) = \eta(x,y)\cdot e^{-\frac{i}{\hbar}Et}D^{\alpha\beta}(E){1\choose\gamma(E)^\beta}
\end{displaymath}
\begin{equation}
\label{psisout}
\times e^{i\alpha\left[k_{zf}+\beta\kappa(E)\right]z}\Theta(\alpha\beta z-a)\Theta(D-|z|)\;\;.
\end{equation}
The solutions of eq. (\ref{tdbdg}) in the normal layer with the constant electric field are
\begin{displaymath}
\Psi_N^{\alpha\beta^\prime}(E;\vec{r},t) = \eta(x,y)\cdot e^{-\frac{i}{\hbar}\left[E+\frac{1}{2}\beta^\prime eV\frac{z}{a}\right]t}C^{\alpha\beta^\prime}(E){\frac{1+\beta^\prime}{2}\choose\frac{1-\beta^\prime}{2}}
\end{displaymath}
\begin{equation}
\label{psin}
\times e^{i\alpha\left[k_{zf}+\frac{\frac{1}{4}eV\frac{z}{a}+\beta^\prime E}{\hbar v_{zf}}\right]z}e^{-i\frac{1}{4}\beta^\prime\Phi_0\frac{z}{a}}\Theta(a-|z|)\;\;.
\end{equation}
They result from an expansion of Airy functions \cite{kue85} in the limit
\begin{equation}
\label{airy}
\frac{\hbar^2k_{zf}^2}{2m^*}\gg |E\pm eV|, \;\;\Delta\;\;.
\end{equation}

The coefficients $D^{\alpha\beta}(E)$ and $C^{\alpha\beta^\prime}(E)$ are determined by matching the superposition 
\begin{equation}
\label{psissuper}
\Psi_S^\alpha(\vec{r},t) = \sum_\beta\int\limits_{-\infty}^\infty \mbox{d}E \left\{\Psi_{S,in}^{\alpha\beta}(E;\vec{r},t)+\Psi_{S,out}^{\alpha\beta}(E;\vec{r},t)\right\}\;\;,
\end{equation}
formed from the solutions (\ref{psisin}) and (\ref{psisout}), to the superposition
\begin{equation}
\label{psinsuper}
\Psi_N^\alpha(\vec{r},t) = \sum_{\beta^\prime}\int\limits_{-\infty}^\infty \mbox{d}E \;\Psi_N^{\alpha\beta^\prime}(E;\vec{r},t)\;\;,
\end{equation}
formed from the solutions (\ref{psin}),
at the NS interfaces at $z=\pm a$:
\begin{equation}
\label{match}
\left.\Psi_S^\alpha(\vec{r},t)\right|_{z=\pm a} = \left.\Psi_N^\alpha(\vec{r},t)\right|_{z=\pm a}\;\;.
\end{equation}
These matching conditions correspond to the ones used in Refs. \cite{kue90,kue85}. 

The integrals over $E$ also contain the solutions which evolve from negative energies $-|E|$. The corresponding wavefunctions
\begin{displaymath}
\Psi_{S,in}^{\alpha\beta}(-|E|;\vec{r},t) \equiv {u_{S,in}^{\alpha\beta}(-|E|;\vec{r},t)\choose v_{S,in}^{\alpha\beta}(-|E|;\vec{r},t)}
\end{displaymath}
are obtained from the ones of positive energy $+|E|$, 
\begin{displaymath}
\Psi_{S,in}^{\alpha\beta}(+|E|;\vec{r},t) \equiv {u_{S,in}^{\alpha\beta}(+|E|;\vec{r},t)\choose v_{S,in}^{\alpha\beta}(+|E|;\vec{r},t)}\;\;,
\end{displaymath}
by the relation 
\begin{equation}
\label{relation}
{u_{S,in}^{\alpha\beta}(-|E|;\vec{r},t)\choose v_{S,in}^{\alpha\beta}(-|E|;\vec{r},t)} = {v_{S,in}^{-\alpha-\beta\;\displaystyle *}(+|E|;\vec{r},t)\choose -u_{S,in}^{-\alpha-\beta\;\displaystyle *}(+|E|;\vec{r},t)}\;\;.
\end{equation}
%
%
%
%
The negative energy states are the ones occupied in the ground state of the system.
\pagebreak

With the wavefunctions (\ref{psisin}), (\ref{psisout}), and (\ref{psin}) the matching conditions (\ref{match}) become:
\smallskip
\begin{displaymath}
\sum_\beta\int\limits_{-\infty}^\infty\!\mbox{d}E e^{-\frac{i}{\hbar}Et}{1\choose\gamma(E)^\beta}e^{i\alpha\left[k_{zf}+\beta\kappa(E)\right]z}\Theta(D-|z|)
\end{displaymath}
\smallskip
\begin{displaymath}
\times\!\left.\left\{N^\beta(E)\Theta(-\alpha\beta z\!-\!a)+D^{\alpha\beta}(E)\Theta(\alpha\beta z\!-\!a)\right\}\right|_{z=\pm a}
\end{displaymath}
\smallskip
\begin{displaymath}
=\sum_{\beta^\prime}\int\limits_{-\infty}^\infty\!\mbox{d}E e^{-\frac{i}{\hbar}\left[E+\frac{1}{2}\beta^\prime eV\frac{z}{a}\right]t}C^{\alpha\beta^\prime}(E){\frac{1+\beta^\prime}{2}\choose\frac{1-\beta^\prime}{2}}
\end{displaymath}
\smallskip
\begin{equation}
\label{fullmatch}
\times e^{i\alpha\left[k_{zf}+\frac{\frac{1}{4}eV\frac{z}{a}+\beta^\prime E}{\hbar v_{zf}}\right]z}\left.e^{-i\frac{1}{4}\beta^\prime\Phi_0\frac{z}{a}}\Theta(a-|z|)\right|_{z=\pm a}\;\;.
\end{equation}
We do not need to know the coefficients $D^{\alpha\beta}(E)$ explicitly, because we calculate the current density in the normal layer of the junction. After straightforward but tedious algebra one finds from eq. (\ref{fullmatch}):
\smallskip
\begin{equation}
\label{CE}
C^{\alpha\beta^\prime}(E) = \sum_{n=0}^\infty\sum_{\beta}c_n^{\alpha\beta\beta^\prime}\left(E-[2n+1-\frac{1}{2}\beta\beta^\prime]\alpha eV\right)
\end{equation}
with
\smallskip
\begin{displaymath}
c_n^{\alpha\beta\beta^\prime}(E) = c_0^{\alpha\beta\beta^\prime}(E) e^{i\left[\frac{4n\left(E-\frac{1}{2}\alpha\beta\beta^\prime eV\right)+4(n^2+n)\alpha eV}{\hbar v_{zf}}\right]a} 
\end{displaymath}
\smallskip
\begin{equation}
\label{cnE}
\times e^{-in\alpha\Phi_0}A_n^\alpha\left(E-\frac{1}{2}\alpha\beta\beta^\prime eV\right)\;\;.
\end{equation}
\smallskip
Here
\begin{displaymath}
c_0^{\alpha\beta\beta^\prime}(E) = N^1(E)\left[1-\gamma(E)^2\right]e^{-i\kappa(E)a}e^{-i(2-\beta\beta^\prime)\frac{1}{4}\alpha\Phi_0}
\end{displaymath}
\smallskip
\begin{equation}
\label{c0E}
\times e^{i\left[\frac{(2-\beta\beta^\prime)(E-\frac{1}{2}\alpha\beta\beta^\prime eV)+\frac{3}{4}\alpha eV}{\hbar v_{zf}}\right]a}\gamma(E+\alpha eV)^{\frac{1-\beta\beta^\prime}{2}}\;\;,
\end{equation}
\smallskip
and
\begin{equation}
\label{AE}
A_n^\alpha(E) = \left\{\begin{array}{ll}\prod\limits_{\nu=1}^{2n}\gamma\left(E+[\nu+\frac{1}{2}]\alpha eV\right) & \mbox{for }n>0\\[3mm] \;1 & \mbox{for }n=0\end{array}\right.
\end{equation}
is the multiple Andreev reflection probability amplitude (cf. \cite{kue90}).

$C^{\alpha\beta^\prime}(E)$ consists of the phase--coherent superpositions of all amplitudes $c_n^{\alpha\beta\beta^\prime}\left(E-[2n+1-\frac{1}{2}\beta\beta^\prime]\alpha eV\right)$ which originate from incident quasiparticles described by the wavefunctions $\Psi_{S,in}^{\alpha\beta}\left(E-[2n+1-\frac{1}{2}\beta\beta^\prime]\alpha eV;\vec{r},t\right)$. In other words: The probability amplitude $C^{\alpha\beta^\prime}(E)$ of finding a quasiparticle (characterized by $\alpha$ and $\beta^\prime$) in the\linebreak N layer, with energy in the range between\linebreak $E - eV/2$ and $E + eV/2$, is the sum of the amplitudes $c_n^{\alpha\beta\beta^\prime}\left(E-[2n+1-\frac{1}{2}\beta\beta^\prime]\alpha eV\right)$; $|c_n^{\alpha\beta\beta^\prime}|^2$ gives the probability that a quasiparticle, incident from one of the S--banks and characterized by $\alpha, \beta$, and energy $E - [2n+1-\frac{1}{2}\beta\beta^\prime]\alpha eV$, reaches the energy range between $E - eV/2$ and $E + eV/2$ in the N layer after n Andreev--reflection cycles, being an electron (hole) for $\beta^\prime=+1$ ($-1$).

However, the most convenient way of calculating the current density $\vec{\jmath}$ in the N layer
%
%
%
%
is not the one which uses the wavefunctions with $ C^{\alpha\beta^\prime}(E)$. It is much\linebreak easier to count properly all quasiparticle contributions to $\vec{\jmath}$ of eq. (\ref{cur}) if one uses new N layer wavefunctions $\Psi_{Nk}(\vec{r},t)$. (These are the wavefunctions (\ref{PSIK}) in the\linebreak N layer.)
%
%
%
%
They are obtained by adding up the phase--coherent wavefunction components which, by multiple Andreev reflections, originate in the N layer from\linebreak {\em one particular} incident quasiparticle described by eq. (\ref{psisin}) with $\alpha, \beta$, and energy $E_0$.
%
%
%
%
The sum of these components
%
%
%
%
is given by the right--hand side of eq. (\ref{psin}) if there one sets $E=E_0+[2n+1-\frac{1}{2}\beta\beta^\prime]\alpha eV$ and replaces $C^{\alpha\beta^\prime}(E)$ by the right--hand side of eq. (\ref{CE}) with one change: Instead of summing over the index $\beta$ characterizing the nature (electron--like or hole--like) of the {\em incident} quasiparticle one has to sum over the index $\beta^\prime$ which characterizes the quasiparticle nature (electron or hole) in the N layer. 
%
%
%
%

One can show: i) The new wavefunctions
\begin{displaymath}
\Psi_{\!Nk}(\vec{r},t)\!\equiv\!\eta(x,y)\cdot\!\sum_{n=0}^\infty\sum_{\beta^\prime}\!e^{-\frac{i}{\hbar}\left[E_0+(2n+1-\frac{1}{2}\beta\beta^\prime)\alpha eV\!+\frac{1}{2}\beta^\prime\! eV\!\frac{z}{a}\right]t}
\end{displaymath}
\begin{displaymath}
\times c_n^{\alpha\beta\beta^\prime}(E_0){\frac{1+\beta^\prime}{2}\choose\frac{1-\beta^\prime}{2}}e^{i\alpha\left[k_{zf}+\frac{\frac{1}{4}eV\frac{z}{a}+\beta^\prime\left[E_0+(2n+1-\frac{1}{2}\beta\beta^\prime)\alpha eV\right]}{\hbar v_{zf}}\right]z}
\end{displaymath}
\begin{equation}
\label{psinenull}
\times e^{-i\frac{1}{4}\beta^\prime\Phi_0\frac{z}{a}}\Theta(a-|z|)\;\;
\end{equation}
match to the solutions in the S layers. ii) In the limit of vanishing voltage, $V\to 0$, the absolute squares of the electron and hole components of $\Psi_{Nk}$ exhibit the resonances of the virtually bound states one has in the stationary situation \cite{gun94}.
%
%
%
%

\vspace{1em}
\centerline{\bf 4. Current--Voltage Characteristics}\vspace{1em}

In the Bogoliubov--de Gennes--formalism the current density in the N layer is \cite{kue69}:
\begin{displaymath}
\vec{\jmath}\,(\vec{r},t)=-\frac{e}{m^*}\sum_k\left\{f_ku_{Nk}^{\displaystyle *}(\vec{r},t)\vec{p}_eu_{Nk}(\vec{r},t)\right.
\end{displaymath}
\begin{equation}
\label{cur}
+\left.\left[1-f_k\right]v_{Nk}(\vec{r},t)\vec{p}_ev_{Nk}^{\displaystyle *}(\vec{r},t)+c.c.\right\}\;\;,
\end{equation}
where the sum over $k\equiv\{s;k_y;(\,E\!\!\equiv\!\! E_0,\;\alpha,\beta)\}$, characterizing the incident quasiparticles, only counts the states with positive energy $E\ge\Delta$ and one spin direction; $f_k=\left(\exp(\frac{E}{k_BT})+1\right)^{-1}$. Applying the momentum operator $\vec{p}_e=\frac{\hbar}{i}\vec{\nabla}+\vec{e}_z\frac{\hbar}{4a}\Phi(t)$ to the wavefunctions\pagebreak\  given by eq. (\ref{psinenull}), averaging over the channel thickness $L_x$, and defining
\begin{displaymath}
\vec{J}^{\;(\alpha\beta\beta^\prime)}(E;z,t)\equiv\vec{e}_z\frac{e}{L_x\pi^2\hbar^2}\frac{1}{\sqrt{k_F^2-k_\varrho^2+\beta\frac{2m^*}{\hbar^2}\sqrt{E^2\!-\!\Delta^2}}}
\end{displaymath}
\begin{displaymath}
\times\frac{E}{\sqrt{E^2-\Delta^2}}\frac{\left[1-\gamma(E)^2\right]^2}{1+\gamma(E)^2}\left|\gamma(E+\alpha eV)^{\frac{1-\beta\beta^\prime}{2}}\right|^2
\end{displaymath}
\begin{displaymath}
\times\mbox{ Re }\left\{\sum_{n,n'=0}^\infty e^{-\frac{i}{\hbar}\alpha 2(n-n')eVt}e^{i\beta^\prime\frac{2(n-n')eVm^*}{\hbar^2\sqrt{k_F^2-k_\varrho^2}}z}e^{-i\alpha(n-n')\Phi_0}\right.
\end{displaymath}
\begin{displaymath}
\times e^{i\left[\frac{4(n-n')\left[E+(n+n'+1-\frac{1}{2}\beta\beta^\prime)\alpha eV\right]}{\hbar^2\sqrt{k_F^2-k_\varrho^2}}\right]am^*}A_n^\alpha(E-\frac{1}{2}\alpha\beta\beta^\prime eV)
\end{displaymath}
\begin{displaymath}
\times A_{n'}^{\alpha\displaystyle *}(E-\frac{1}{2}\alpha\beta\beta^\prime)\cdot\alpha\Bigg[\hbar\sqrt{k_F^2-k_\varrho^2}
\end{displaymath}
\begin{equation}
\label{jabb}
+\left.\left.\frac{\frac{eVz}{2a}+\beta^\prime\left[E+(2n+1-\frac{1}{2}\beta\beta^\prime)\alpha eV\right]}{\frac{\hbar}{m^*}\sqrt{k_F^2-k_\varrho^2}}\right]\Theta(a-|z|)\right\}\;\;,
\end{equation}
we obtain the current density in the N layer as
\begin{displaymath}
\vec{\jmath}\,(z,t)=-\sum_{\alpha\beta}\int\limits_\Delta^{\hbar\omega_D}\mbox{d}E\sum_{s=1}^2\int\limits_0^{k_{y,\max}}\mbox{d}k_y\left(f_k\vec{J}^{\;(\alpha\beta\, 1)}(E;\vec{r},t)\right.
\end{displaymath}
\begin{equation}
\label{jzt}
-\left.\left[1-f_k\right]\vec{J}^{\;(\alpha\beta\, -1)}(E;\vec{r},t)\right)\;\;.
\end{equation}
As usual we integrate over all energies up to the Debye energy $\hbar\omega_D$;\newline $k_{y,\max}\equiv\left(k_F^2-\frac{2m^*}{\hbar^2}\left(\sqrt{E^2-\Delta^2}+\Delta\right)-\left(\frac{\pi}{L_x}\right)^2s^2\right)^{1/2}$.

The time average of $\vec{\jmath}\,(z,t)$ is
\begin{equation}
\label{tmitt}
\vec{\jmath}\,(z)=\frac{1}{T_M}\int\limits_0^{T_M}\mbox{d}t \vec{\jmath}\,(z,t)
\end{equation}
with $T_M\rightarrow\infty$. In numerical calculations we choose $T_M=1$ s, which is very much larger than all characteristic time scales.

The time--dependent current density $\vec{\jmath}(0,t)$ in the center of the 
junction is computed by evaluating  numerically  
equation (\ref{jzt})  for the voltage--bias case, i.e. $\Phi_0 = 0$. 
We assume  a 
proximity--induced pair potential\linebreak $\Delta = 0.3$ meV. The other parameters are 
indicated in Section 2. The result is shown in Fig. \ref{acj2D}. For comparison
Fig. \ref{acj3D} shows the time--dependent current density of a 
three--dimensional 
SNS junction with the same pair potential but translational invariance in $x$-- and $y$--directions \cite{jacobs97}. Its basis is the 3D current density equation which one  obtains from   eq. (\ref{jzt}) essentially by  changing the sum over the indices s=1 and 2 of the two occupied subbands into an integral 
over 
continous\linebreak $k_x$--values, and by replacing the effective mass $m^*$ by the 
\begin{center}
\includegraphics[width=8cm]{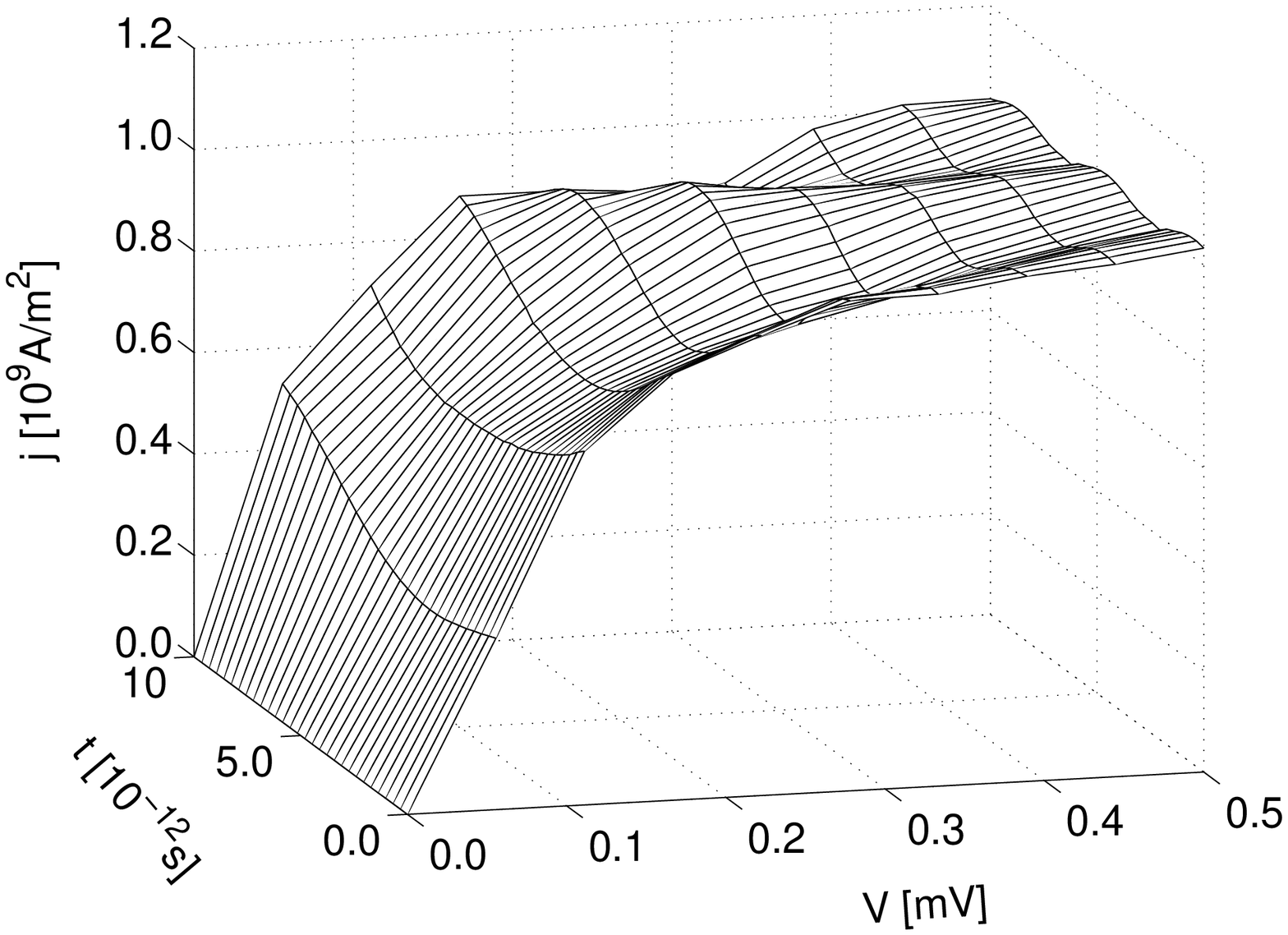}\\[-7mm]
\end{center}
\vspace{-2em}
\begin{fig}The time-- and voltage--dependent current density at temperature $T=2.2$ K in the quasi--twodimensional  electron gas of an SSmS junction with the proximity--induced pair potential of Fig. \ref{pp}. The length of the Sm--layer is $2a=500$ nm, and the electron density is 5$\cdot10^{18}$cm$^{-3}$. \label{acj2D}
\end{fig}
\begin{center}
\includegraphics[width=8cm]{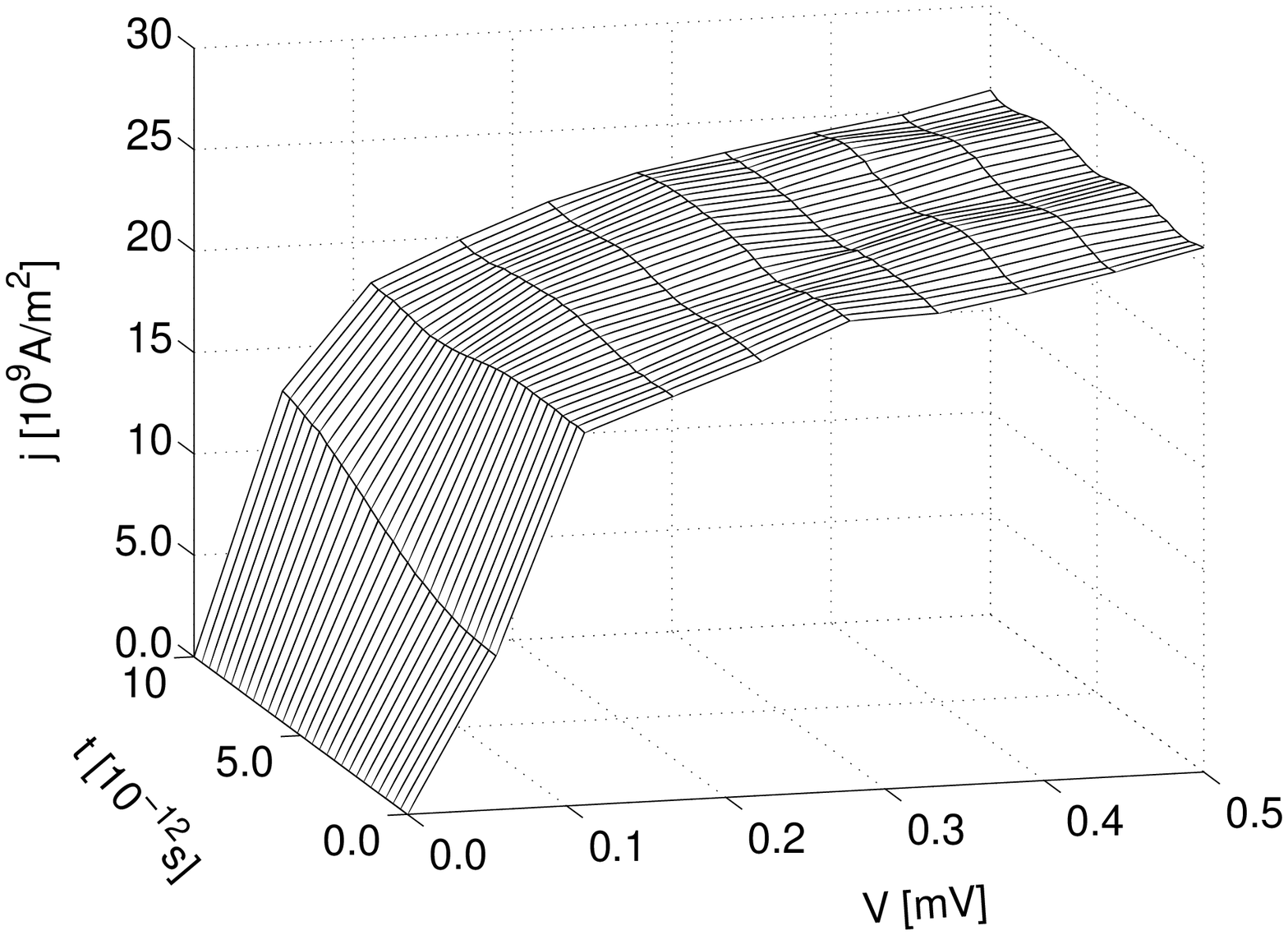}\\[-7mm]
\end{center}
\vspace{-2em}
\begin{fig}The time-- and voltage--dependent current density at temperature $T=2.2$ K in a threedimensional SNS junction. The length of the N layer is $2a=500$ nm, and the Fermi energy is 0.21 eV. \label{acj3D}
\end{fig}
free electron mass $m_0$. Not surprisingly, there are only quantitative differences: The fundamental mechanism of multiple Andreev reflections, responsible for the charge transport in mesoscopic weak links, works the same way in three-- and quasi--twodimensional electron gases. The amplitudes of the current density oscillations 
over time  are small. Therefore, the time--averaged current densities $\vec{\jmath}(0)$, computed from  eq. (\ref{tmitt}) and presented in Fig. \ref{dcj2D}, result in current--voltage characteristics (CVCs) which do not differ much from the CVCs in Fig. \ref{acj2D} at any fixed time. (In order to save computer time, the CVCs of Figs. \ref{acj2D} and \ref{acj3D} have been calculated with less numerical accuracy than those of  Fig. \ref{dcj2D}. Therefore, the latter are 
 smoother  than the former. Also, the low--voltage slope in Fig. \ref{acj2D} is not as steep as in Fig. \ref{dcj2D} because only four, instead of 36, Andreev reflection cycles have been taken into account in its computation.) 
\begin{center}
\includegraphics[width=8cm]{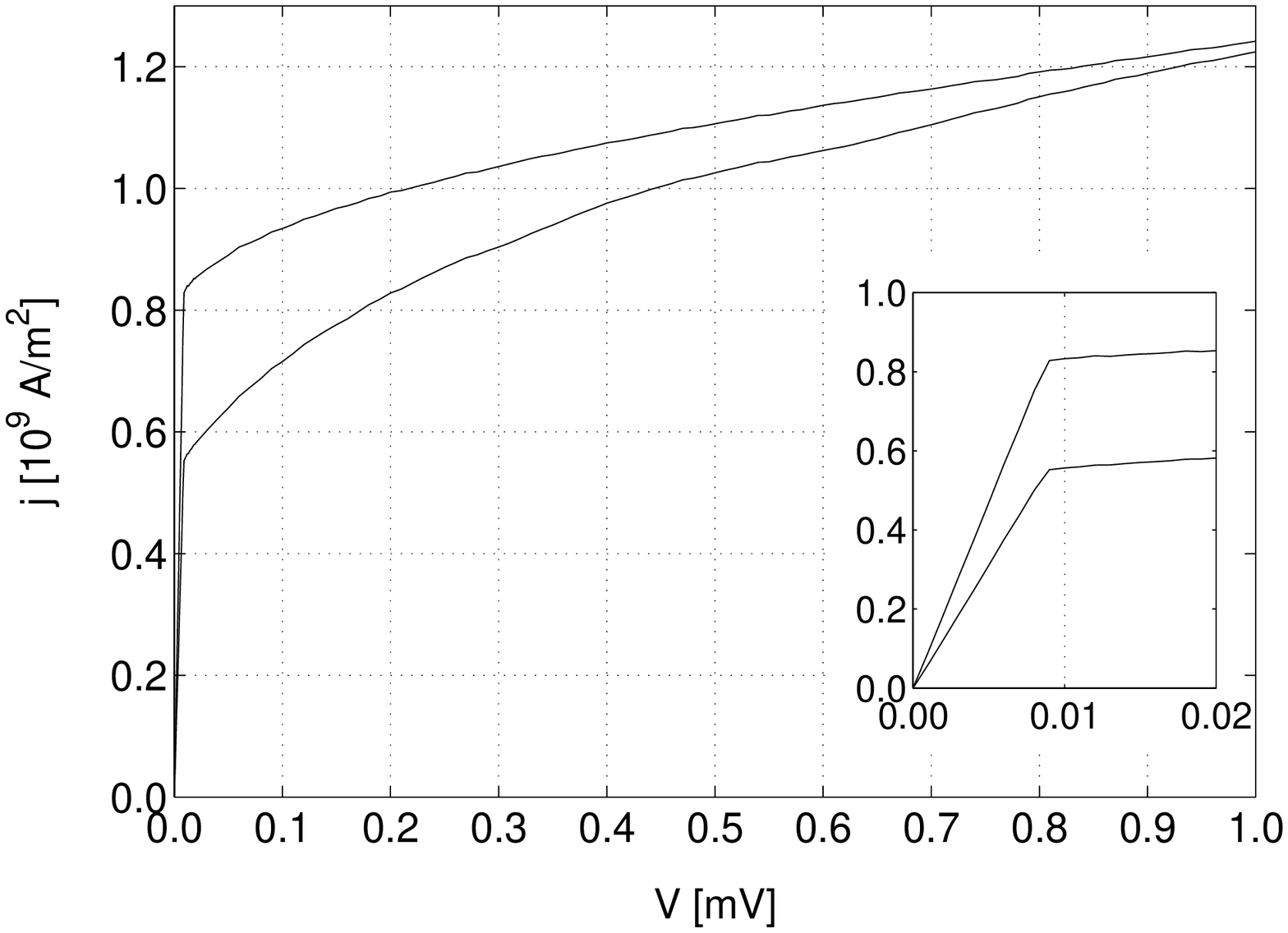}\\[-7mm]
\end{center}
\vspace{-1.3em}
\begin{fig}Current--voltage characteristics of a
Q2D SSmS junction at $T=2.2$ K (lower curve) and\hspace{3cm} $T=0$ K (upper curve), same parameters as in\hspace{2cm} Fig. \ref{acj2D}. The inset shows the steep rise of the current density with small voltages; again, the lower curve is for the higher temperature. \label{dcj2D}
\end{fig}

\vspace{1em}

\centerline{\bf 5. Discussion}\vspace{1em}

The computed current densities in Figs. \ref{acj2D} -- \ref{dcj2D}  
agree with the 
experimental findings of  Thomas et al.
\cite{tho98} in the following qualitative aspects: i) There are 
voltage--dependent
current--oscillations in time. The period is the inverse of the Josephson 
frequency $\nu = 2eV/h$: The dominant contributions of the 
multiple Andreev reflections to the  alternating (Josephson) current come from the terms with 
$n - n' = 1$ in eq. (\ref{jabb}).    ii) The CVCs of the time--averaged current densities exhibit what Thomas et al. call the ``corner current''. This  current value
separates the steep, nearly linear increase of the current with very small voltages (also called the ``foot'' in prior investigations of non--equilibrium effects in weak links \cite{kue90}) from a wide range of slow current increase 
with higher voltages. The physical mechanism responsible for the drastic 
change of current growth with voltage is the decreasing  number of 
relevant Andreev reflection cycles
in the energy range between $-\Delta $ and $+\Delta $  with increasing 
voltages above  the voltage  of the  ``corner current''.  
iii) The zero--bias differential resistance $dV/dI|_{V=0}$ increases (and the 
current decreases) with temperature. 
These points of qualitative agreement
between experiment and theory support the view of Thomas et al. that multiple
Andreev reflections are responsible for the observed nonequilibrium effects
\cite{tho98}.
The characteristic shape of the CVC described in  ii) has been also observed 
for a single SNS junction by Kroemer et al. \cite{kro94a}. There is even
satisfactory {\em quantitative} agreement between the magnitude of the 
experimental current densities and voltages in Fig. 2 of Ref. \cite{kro94a} and the computed ones in our Fig. \ref{dcj2D}. 
Not surprisingly, quantitative agreement  with the measurements of Thomas et 
al. performed on 
more than 300 SNS junctions in series \cite{tho98} is not so good: 
Our current densities are at least four times greater than theirs.
For many SNS junctions in series the following approximations of our model may have to be modified: It
a) confines all electrons to the channel, neglecting their penetration into 
the Nb contacts, b) neglects trapping of  electrons by surface states in 
$y = 0$ and $y = L_y$, and c) disregards the small probability of conventional scattering 
from spatial variations of the pair potential \cite{kue77,sip96}, which may 
occasionally disrupt the Andreev reflection cycles \cite{tho98}. 

Two important qualitative features of the experimental CVCs are not reproduced by our model.  The first is the  dependence of the steep  slope of the 
CVCs for vanishing voltages
(the ``apparent activation energy'' in the Arrhenius--plot language 
of Thomas et al.) on the sample width $L_y$. In order to include effects of 
this width in the model, one may have to take into account -- via a phenomenological, $L_y$--dependent relaxation time -- that the phase--coherence of Andreev reflections can be destroyed by the temporary trapping of electrons in surface states. Alternatively, one may also think of inhomogeneities in the current density, caused by the (neglected) magnetic field of the current and the finite Josephson penetration depth.  The required improvements would not change the fundamental character of the model. A significant model change, however, may be 
necessary in order to reproduce the second characteristic  feature of the experimental CVCs not present in our figures: The nearly flat plateau in the CVCs between the ``corner current'' at the end of the low--voltage ``foot''   and the 
linearly rising excess current at high voltages. Such a plateau is absent in all theories where the principal contributions to the Multiple--Andreev--Reflection current are from quasiparticles which evolve from (scattering) states with energies {\em above} the gap. Our CVCs, calculated with solutions of the time--dependent Bogoliubov--de Gennes Equation, and that of Gunsenheimer and Zaikin, obtained with the 
Keldysh technique \cite{gun94a}, are qualitatively quite\linebreak similar and lack the plateau. On the other hand,  CVCs computed with 
 quasiparticle wavefunctions which, at the rate of Andreev reflection 
occurrence, evolve from states with energies {\em below} $\Delta $  show the 
plateau for appropriate inelastic mean free paths \cite{kue90}. 
As discussed in Section 3,
 this should correspond to a situation where the sample is connected  to the 
external energy and particle reservoir by normal conducting current leads, and
where hybrid quasiparticle/collective--mode excitations at energies smaller 
than $\Delta$ exist.   The 
inclusion of the subgap starting states in the theory, also necessary in order to obtain the  direct Josephson current in  the  
 case of current bias, $\Phi_0 \neq 0$, and vanishing voltage, is a subject of work in 
progress. 

\vspace{1em}
\smallskip
\noindent {\bf Acknowledgements} -- 
This work was initiated at the ESPRIT/HCM/PECO--Workshop on ``Novel Aspects of
Superconductivity in Layered Materials, Confined Geo\-metries and Strongly Correlated Electron Systems'' of the {\em Institute for Scientific Interchange}, Foundation Torino. One of us, R. K., gratefully acknowledges the hospitality
extended to the workshop participants by  the Villa Gualino. We thank Dr. Uwe 
Gunsenheimer for discussions and computational advice and Professor Herbert 
Kroemer for detailed information on his work prior to its publication.

\bibliography{artikel,super}

\end{multicols}

\end{document}